\documentclass[]{elsarticle}

\usepackage[utf8]{inputenc}
\usepackage{graphicx}
\usepackage{subfigure}

\usepackage{amssymb}
\usepackage{amsthm}
\usepackage{url}
\usepackage{amsmath}
\usepackage{subfigure}
\usepackage{textcomp}
\usepackage{xcolor}
\usepackage{xcolor}
\usepackage[margin=2.5cm]{geometry}
\usepackage{gensymb}

\graphicspath{{figs/}}

\title{Reactive fungal wearable}

\author[1,*]{Andrew Adamatzky}
\author[1,3]{Anna Nikolaidou}
\author[2]{Antoni Gandia}
\author[1,4]{Alessandro Chiolerio}
\author[1,5]{Mohammad Mahdi Dehshibi}

\address[1]{Unconventional Computing Laboratory, UWE, Bristol, UK}
\address[2]{Mogu S.r.l., Inarzo, Italy}
\address[3]{Department of Architecture, UWE, Bristol, UK}
\address[4]{Center for Sustainable Future Technologies,
Istituto Italiano di Tecnologia, Torino, Italy}
\address[5]{Department of Computer Science, Universitat Oberta de Catalunya, Barcelona, Spain}

\journal{Journal Name}

\begin{document}

\begin{frontmatter}

\begin{abstract}
Smart wearables sense and process information from the user's body and environment and report results of their analysis as electrical signals. Conventional electronic sensors and controllers are commonly, sometimes augmented by recent advances in soft electronics. Organic electronics and bioelectronics, especially with living substrates, offer a great opportunity to incorporate parallel sensing and information processing capabilities of natural systems into future and emerging wearables. Nowadays fungi are emerging as a promising candidate to produce sustainable textiles to be used as ecofriendly biowearables. To assess the sensing potential of fungal wearables we undertook laboratory experiments on electrical response of a hemp fabric colonised by oyster fungi~\emph{Pleurotus ostreatus} to mechanical stretching and stimulation with attractants and repellents. We have shown that it is possible to discern a nature of stimuli from the fungi electrical responses. The results paved a way towards future design of intelligent sensing patches to be used in reactive fungal wearables.
\end{abstract}


\end{frontmatter}

\section{Introduction}
\label{introduction}

Smart wearables are devices that extend the functionality of clothes and gadgets, they are responsive to the wearer and can act as an interface between the wearer and the environment producing a user responsive symbiotic system. The smart wearables have been developed as a result of the convergence between textiles and electronics (e-textiles). They integrate a high level of technology to provide complex functions and an easy operation and maintenance ~\cite{stoppa2014}. They can be divided into three subgroups: 
(1)~passive smart wearables: able to sense the environment/user, 
(2)~active or reactive smart wearables: able to sense the environment/user, and react performing some actions, therefore integrating an actuator, 
(3)~advanced smart wearables: able to sense, react and adapt their behaviour to given circumstances.
Sensors provide means to detect signals, actuators react upon stimuli either autonomously or after commands received from a central processing unit~\cite{langereis2012}. Textile-embedded sensing systems have been developed and commercially exploited in both the biomedical and safety communities~\cite{custodio2012}. Smart wearables have been used to record electrocardiography signals~\cite{coosemans2003}, electromyography signals~\cite{scalisi2015}, electroencephalography signals~\cite{lofhede2012}, temperature~\cite{sibinski2010}, biophotonic sensing~\cite{omenetto2013}, movement~\cite{meyer2006}, oxygen content, salinity, moisture, or contaminants~\cite{coyle2010, zadeh2006}.
Active functionalities might include power generation or storage capabilities~\cite{vatansever2011}, machine to human interface elements~\cite{baurley2004}, radio frequency communication capabilities~\cite{black2007}. Wearable intelligent systems, intrinsically soft and better compliant with extension, deformation and skin stiffness have been developed since a long time ~\cite{rajan2018}.

The electronic wearables cannot self-grow and self-repair. This deficiency limits their application in the field of soft robotics and self-growing robots~\cite{mazzolai2017plant,sadeghi2017toward, del2018toward, sadeghi2020passive}. We can overcome these limits by incorporating living fabric in the smart wearables. One of the solutions, explored by us previously, would be to grow slime mould \emph{P. polycephalum} on a surface of the cloths or a body of a robot~\cite{schubert2015bodymetries}. The slime mould is proven to be a biosensor for the chemical, mechanical and optical stimuli~\cite{whiting2014towards,adamatzky2013slime,adamatzky2013towards}. Despite the sufficient sensorial abilities, the slime mould is rather fragile, highly dependent on environmental conditions and requires particular sources of nutrients. 

Fungi could, however, make a feasible alternative to the slime mould. Fungal materials --- grow substrates colonised with mycelium of filamentous fungi --- are emerging to be robust, reliable and ecologically friendly replacement for conventional building materials and  fabrics~\cite{travaglini2016manufacturing,haneef2017advanced,ross2018method,appels2019fabrication,islam2017morphology,dahmen2017soft,adamatzky2019fungal,chase2019fungal,meyer2020growing,jones2020mycoleather}. Fungi  ``possess almost all the senses used by humans''~\cite{bahn2007sensing}. Fungi sense light, chemicals, gases, gravity and electric fields. Fungi show a pronounced response to changes in a substrate pH~\cite{van2002arbuscular}, mechanical stimulation~\cite{kung2005possible}, toxic metals~\cite{fomina2000negative}, CO$_2$~\cite{bahn2006co2}, stress hormones~\cite{howitz2008xenohormesis}. Thus, wearables made of or incorporating fabric colonised by fungi might act as a large distributed sensorial network. Fungi is known to respond to chemical and physical stimuli by changing patterns of its electrical activity~\cite{olsson1995action,adamatzky2018spiking,adamatzky2020fungal} and electrical properties~\cite{beasley2020fungal}. This feature would allow to interface fungal wearables with conventional electronics. In view of their extension and interconnectivity, fungal networks represent certainly a sustainable infrastructure-forming substrate, able to wire loci separated by considerable space. Moreover, there are indications that mycelium networks not just sense the external stimuli but also process information, and that there is feasible opportunity to convert fungal responses into Boolean circuits, thus making fungal wearables parallel biological processing networks~\cite{adamatzky2020boolean}. Previously conducted experiments on sensorial properties of fungi have been using experimental laboratory setups where substrates colonised by fungi have been kept in `comfortable' conditions of closed containers with preserved humidity \cite{dehshibi2020electrical}. To assess the feasibility of a fungal wearable prototype in the real world we conducted experiments with a thin hemp-mycelium composite fabric incorporated on a t-shirt wore by a mannequin. 

The paper is structured as follows. We introduce experimental techniques in Sect.~\ref{methods}. Section~\ref{results} analyses the fungal response to stimulation with chemo-attractants, chemo-repellents and mechanical stretching. In Sect.~\ref{discussion} we talk about mechanisms of the fungal wearable response and propose directions for further studies.

\section{Methods}
\label{methods}

\begin{figure}[!tbp]
    \centering
        \subfigure[]{\includegraphics[width=0.75\textwidth, height=0.7\linewidth]{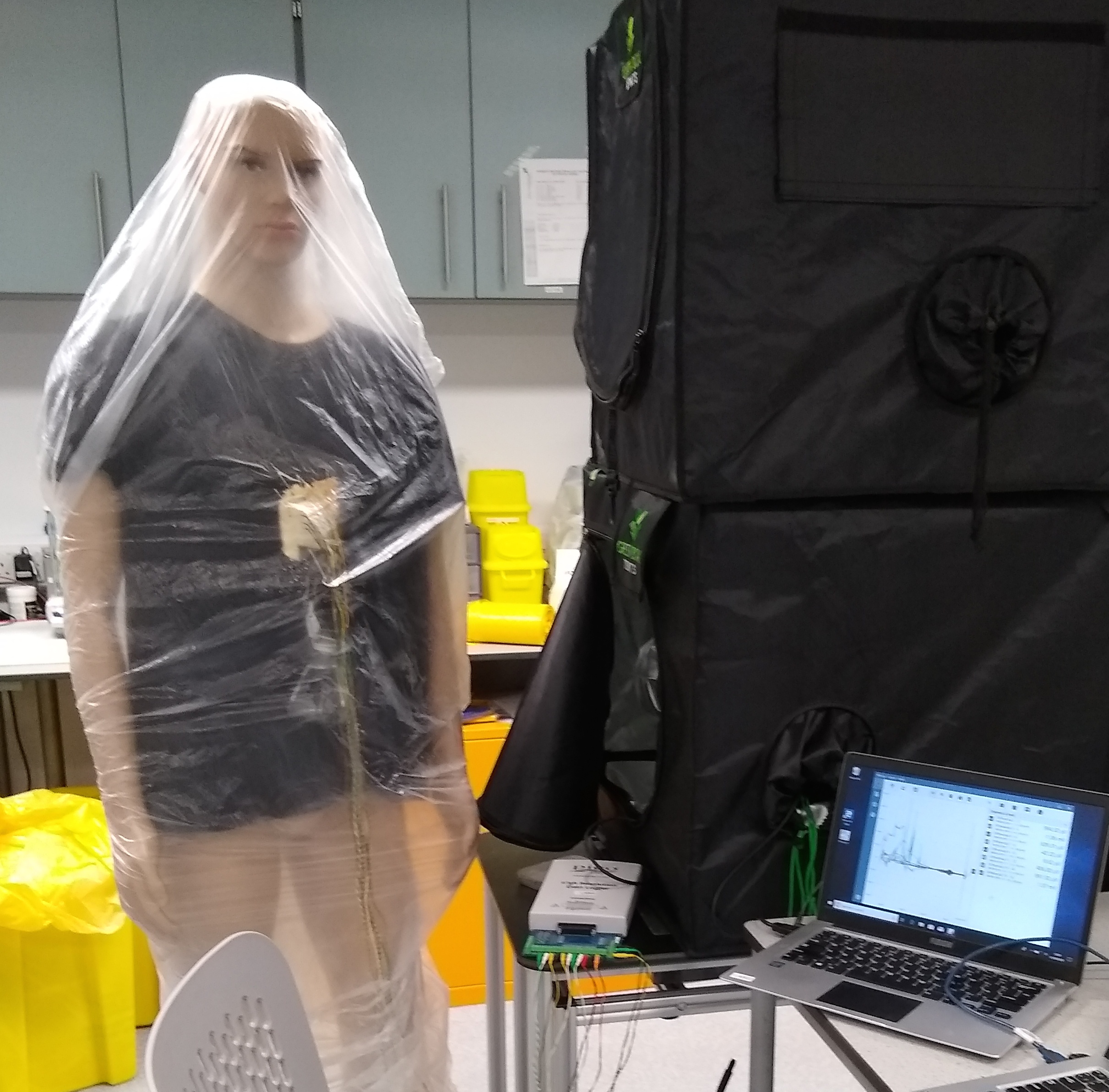}\label{fig:mannequin}}
            \subfigure[]{
    \includegraphics[width=0.28\textwidth, height=0.3\linewidth]{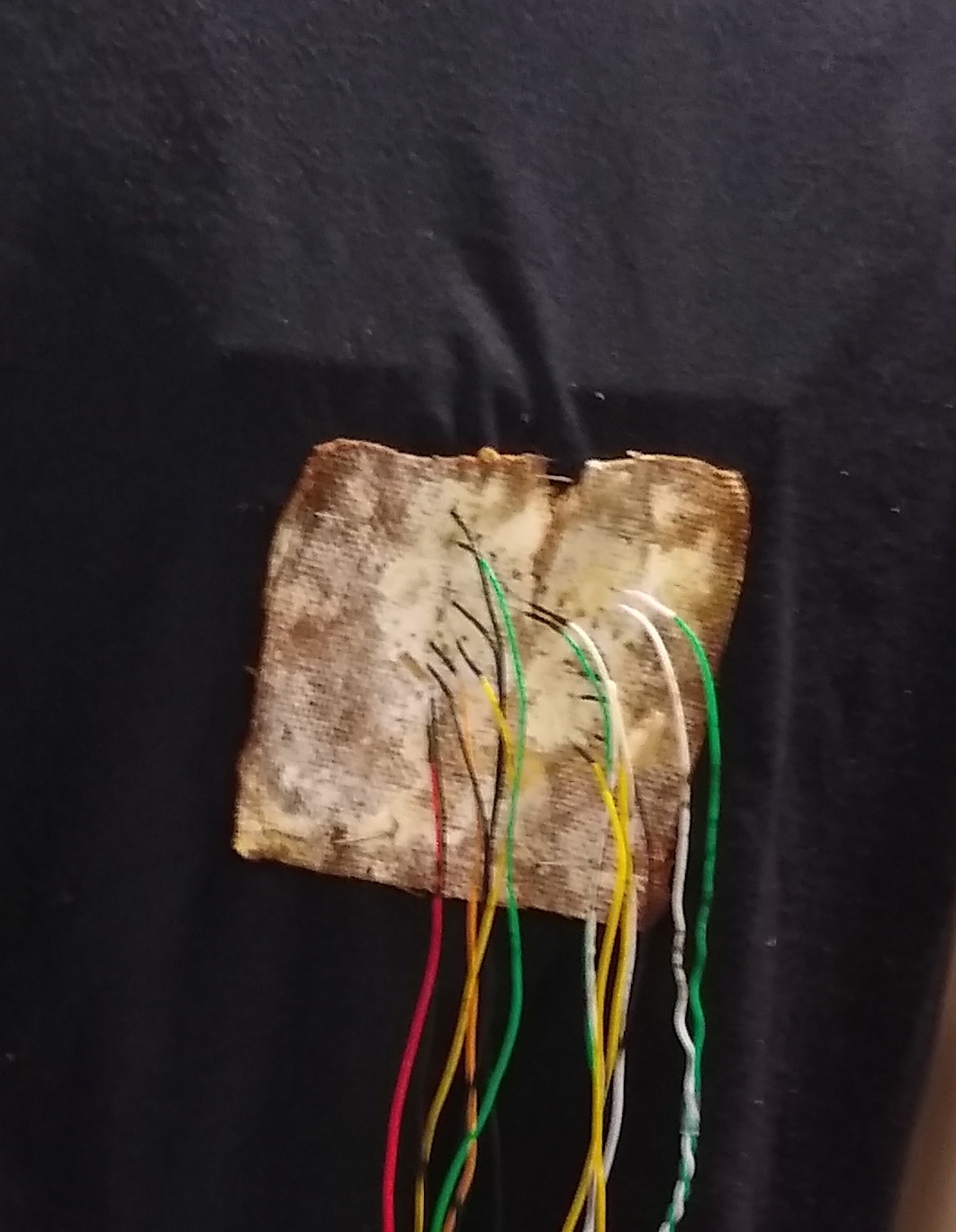}\label{fig:closeup}}
    \subfigure[]{
    \includegraphics[width=0.45\textwidth, height=0.3\linewidth]{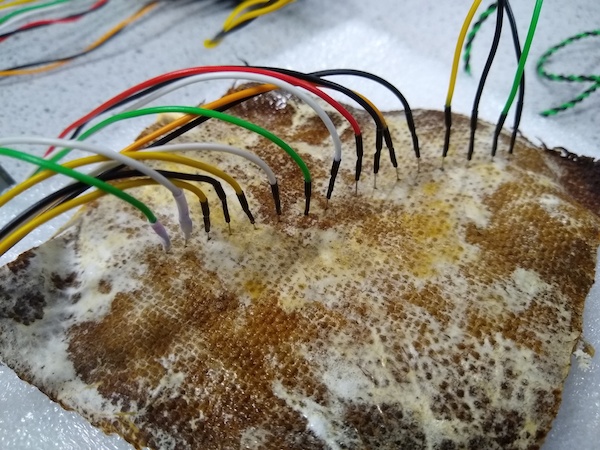}\label{fig:electrodes}}
    \caption{Experimental setup. 
    (a)~Overall view of the experimental setup.
    (b)~Close up of the fungal wearable incorporated into real cloth.
    (c)~Exemplar locations of electrodes.  }
    \label{fig:setup}
\end{figure}

A commercial strain of the fungus \emph{P. ostreatus} (Mogu's collection code 21-18) was cultured on a hemp bedding substrate in plastic boxes c. $35 \times 20~{\text cm}^2$ in darkness at ambient room temperature c.~22\textsuperscript{o}C.

A hemp substrate well colonised by the fungus was spread on rectangular fragments, c. $12 \times12~\text{cm}^2$, of moisturised hemp fabric. When the fragments were colonised, as visualised by white mycelial growth on surface, they were used for experiments. The colonised fabric was attached to a cloth, which in turn was placed on a mannequin (Fig.~\ref{fig:mannequin}). The mannequin was covered by a plastic sheet to prevent a quick decrease of moisture in the fungal fabric. The fabric was sprayed with distilled water once per two days. The humidity of the fungal fabric was 70\%-80\% (MerlinLaser Protimeter, UK). The experiments were conducted in a room with ambient temperature 21\textsuperscript{o}C and illumination 30-150 LUX (ISO-Tech ILM 1332A).

Electrical activity of the colonised fabric was recorded using pairs of iridium-coated stainless steel sub-dermal needle electrodes (Spes Medica S.r.l., Italy), with twisted cables and  ADC-24 (Pico Technology, UK) high-resolution data logger with a 24-bit A/D converter, galvanic isolation and software-selectable sample rates all contribute to a superior noise-free resolution. To keep electrodes stable we have been placing a polyurethane pad under the fabric. The pairs of electrodes were pierced through the fabric and into the polyurethane pad (Figs~\ref{fig:closeup} and \ref{fig:electrodes}). 
We recorded electrical activity one sample per second, where the minimum and maximum logging times were 60.04 and 93.45 hours, respectively. During the recording, the logger has been doing as many measurements as possible (typically up to 600 per second) and saving the average value. We set the acquisition voltage range to 156~mV with an offset accuracy of 9~$\mu$V at 1~Hz to maintain a gain error of 0.1\%. Each electrode pair was considered independently with the noise-free resolution of 17 bits and conversion time of 60~ms. Each pair of electrodes, further called a Channel (Ch), reported a difference of the electrical potential between the electrodes. Distance between electrodes was 1-2~cm. In each trial, we recorded 5–8 electrode pairs (Ch) simultaneously. We stimulated the fungus with 96\% ethanol, malt extract powder (Sigma Aldrich, UK) dissolved in  distilled water, dextrose (Ritchie Products Ltd, UK) and by attaching weights (using foldback clips) to the hemp pads.

\section{Results}
\label{results}

\begin{figure}[!tbp]
    \centering
    \subfigure[]{\includegraphics[width=0.49\textwidth]{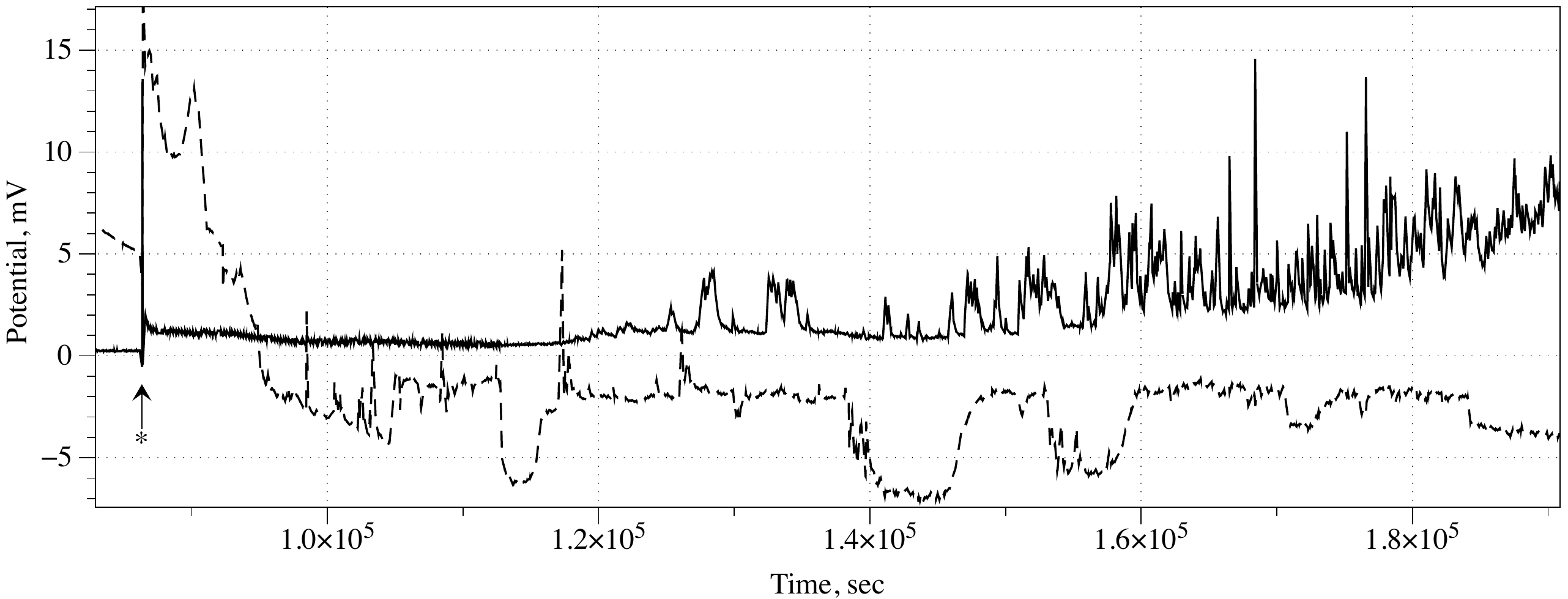}\label{fig:malt1}}
    \subfigure[]{\includegraphics[width=0.49\textwidth]{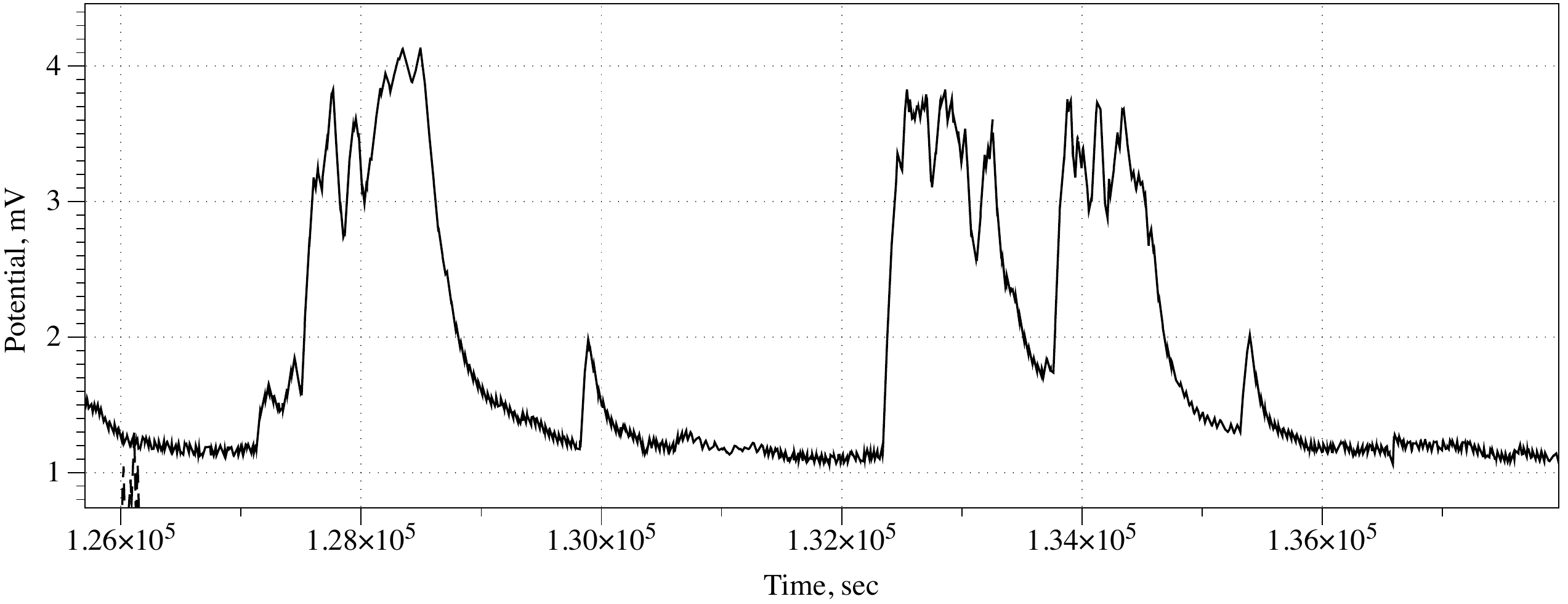}\label{fig:trainLowFreq}}
    \subfigure[]{\includegraphics[width=0.49\textwidth]{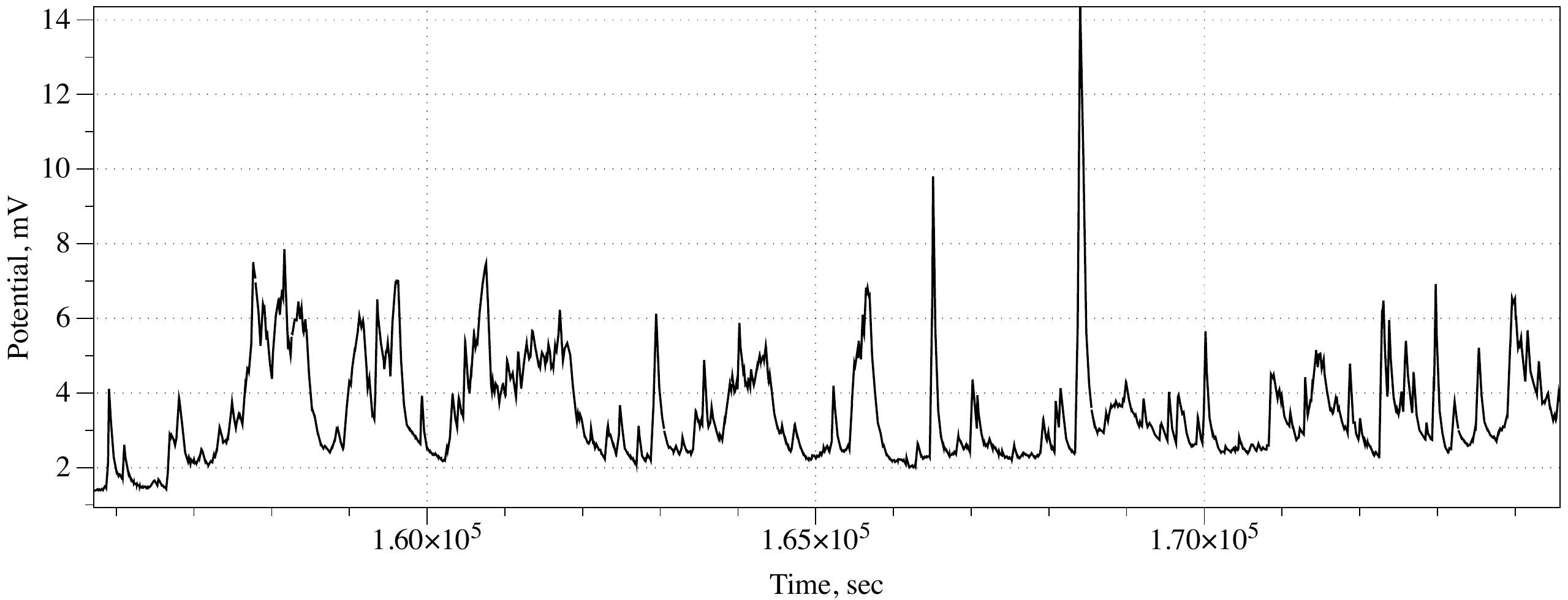}\label{fig:trainHighFreq}}
    \subfigure[]{\includegraphics[width=0.49\textwidth]{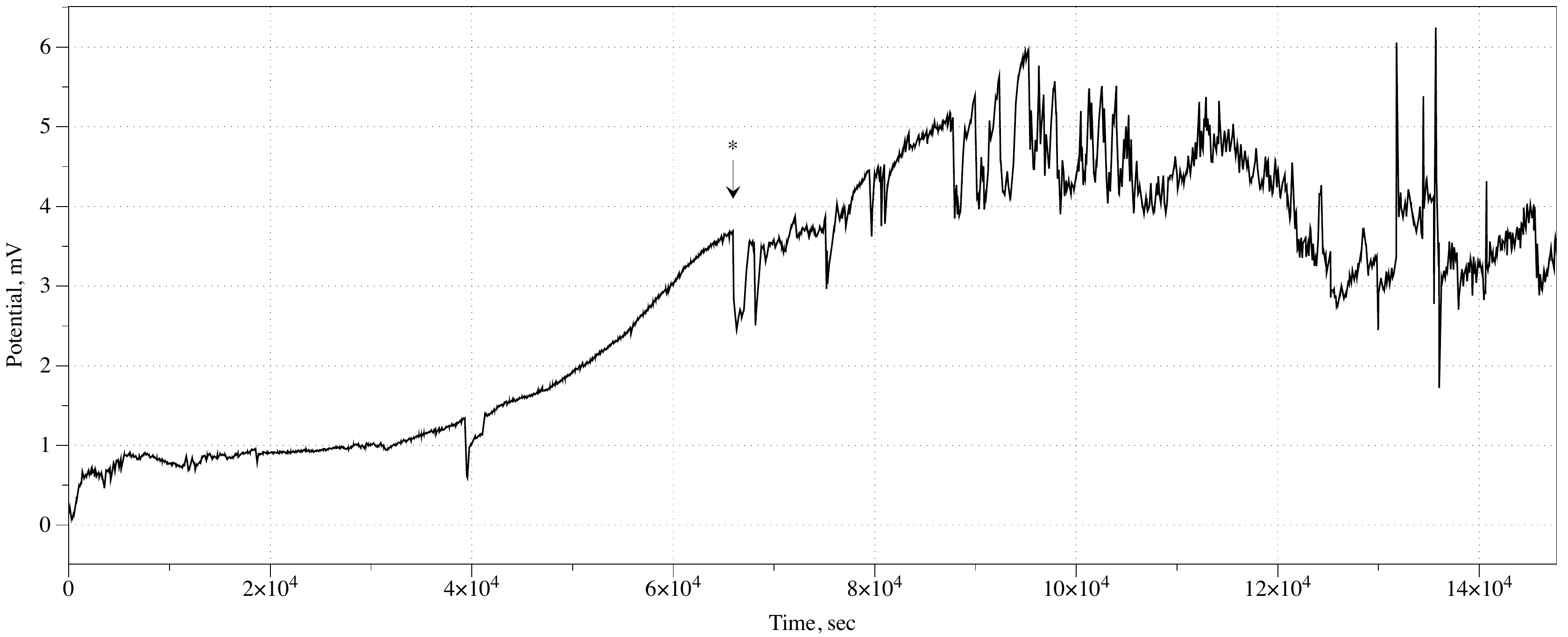}\label{fig:dextrose}}
    \subfigure[]{\includegraphics[width=0.4\textwidth]{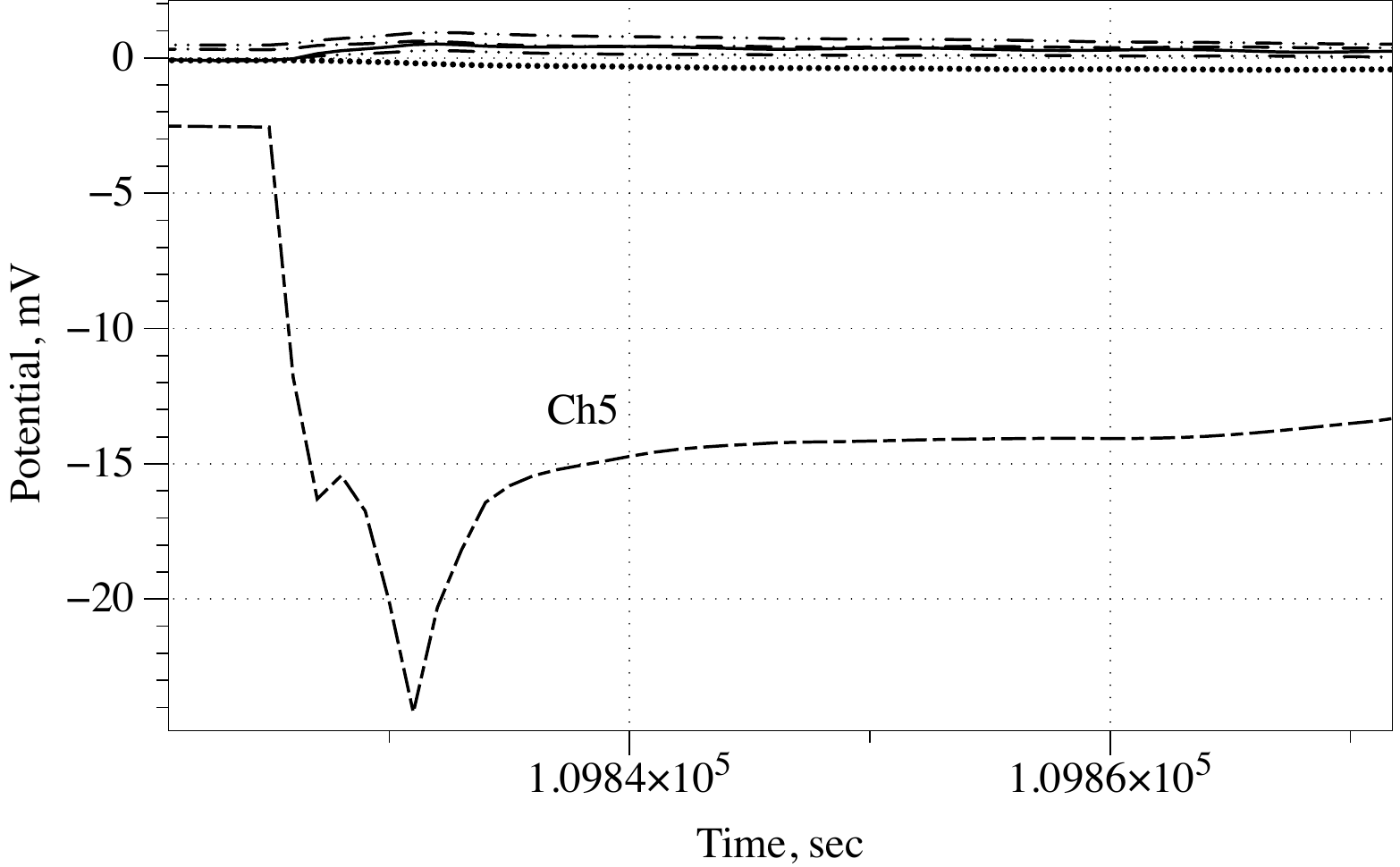}\label{fig:ethanolCh5}}
    \subfigure[]{\includegraphics[width=0.4\textwidth]{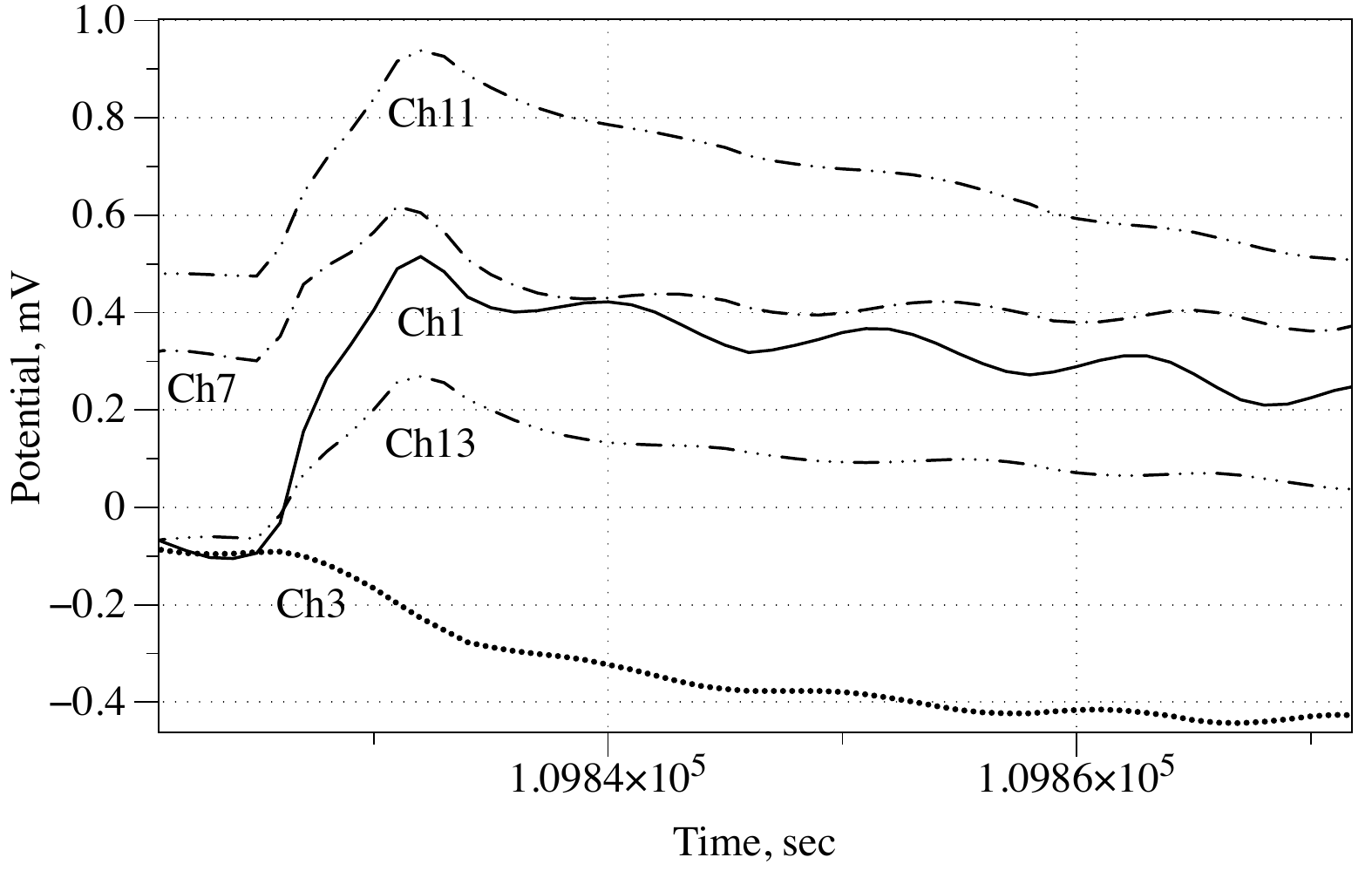}\label{fig:ethanolA}}
    \caption{(a)~Response to application of malt extract. The moment of malt extract application is shown by asterisk. 
    (b)~Low frequency trains of spikes. 
    (c)~High frequency trains of spikes.
    (d)~Response to application of dextrose. The moment of malt extract application is shown by asterisk. 
    (ef)~Response to stimulation with ethanol. An overall response is shown in (e)~with some channels zoomed in~(f).}
\end{figure}

A response of the fungal wearable to  a chemo-attractant was studied using malt extract and dextrose. An exemplar response to application of malt extract is shown in Fig.~\ref{fig:malt1}. The immediate, first 3~hr, response is manifested in the spikes up to 15~mV and duration up to 140~min, and is due to a sudden increase in humidity of the substrate. Further response is attributed to fungi sensing malt extract as a chemo-attractant and a source of nutrients. The onset of the response is characterised by low frequency trains of spikes (Fig.~\ref{fig:trainLowFreq}). There are 3-4 spikes, with amplitude over 2~mV, in each train. Average distance between spikes in each train is 291~sec, $\sigma=90$, average spike width is 273~sec,  $\sigma=110$, average spike amplitude 2.6~mV, $\sigma=0.15$. Average duration of a train is 31~min, $\sigma=3$, a distance is up to one hour.  Typically, a frequency of spike trains increases with time, a distance between trains decreases to 15~min in average, $\sigma$=5 (Fig.~\ref{fig:trainHighFreq}). Average amplitude of spike trains is 4.6~mV, $\sigma$=2.5.

Results of experiments with malt extract echo in the experiments with application of dextrose (Fig.~\ref{fig:dextrose}). The fungi show low frequency oscillatory activity before stimulation: average distance between spikes is 197~min, $\sigma=13.9$, average amplitude 0.3~mV, $\sigma=0.2$. In first 5 hours after the dextrose application the frequency of spikes substantially increases: average distance between spikes becomes 22~min, $\sigma=13$ and amplitude increases to average
0.43~mV, $\sigma=0.27$. In next 5 hours average amplitude of spikes increases to 1.3~mV, $\sigma=0.35$, and distance between spikes 
20~min, $\sigma=7$.

To assess a response to chemo-attractants we used ethanol. We applied 1~ml of 96~\% ethanol on the colonised fabric near loci of Ch5. The response on one of the channel close to the application loci is shown in Fig.~\ref{fig:ethanolCh5}.  We observed a drop by nearly 8~mV followed by further drop of the electrical potential by nearly 8~mV. The time to the peak of the response is c.~7~sec. The drop in potential followed by repolarisation phase, which lasted c.~14~sec. The potential remained c.~11~mV lower than that before stimulation. The response on channels further from the application loci is less pronounced, as seen in Fig.~\ref{fig:ethanolA}. The spikes of electrical potential recorded on the channels have the following amplitudes: 0.65~mV on Ch1, 0.34~mV on Ch3, 0.31~mV on Ch7, 0.5~mV on Ch11, and 0.3~mV on Ch13, where Ch1 is the closest to the application loci and Ch13 is the farthest one.  

\begin{figure}[!tbp]
    \centering
    \subfigure[]{\includegraphics[width=0.49\textwidth]{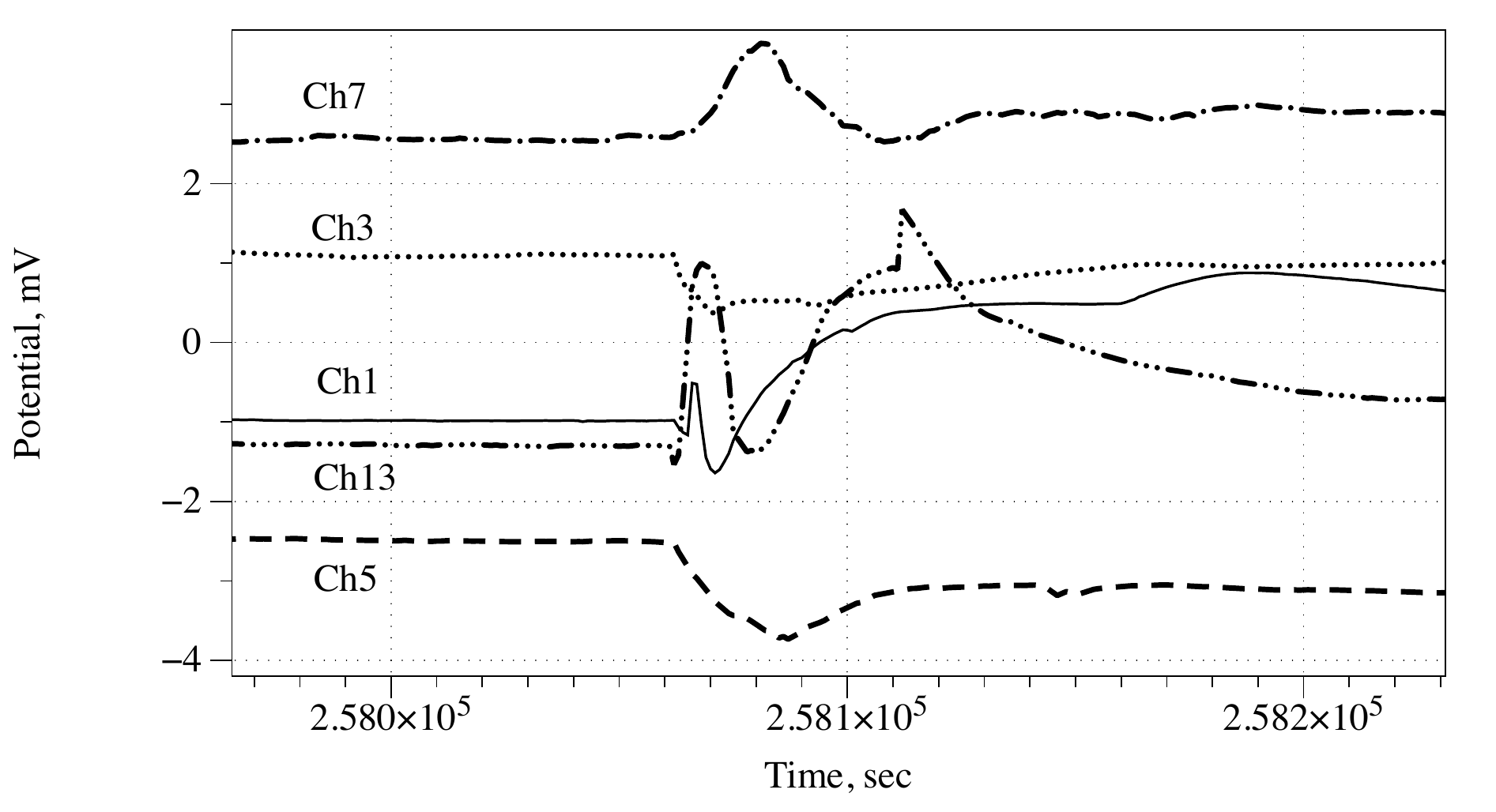}\label{fig:lab01_w_50_g_applied}}
    \subfigure[]{\includegraphics[width=0.49\textwidth]{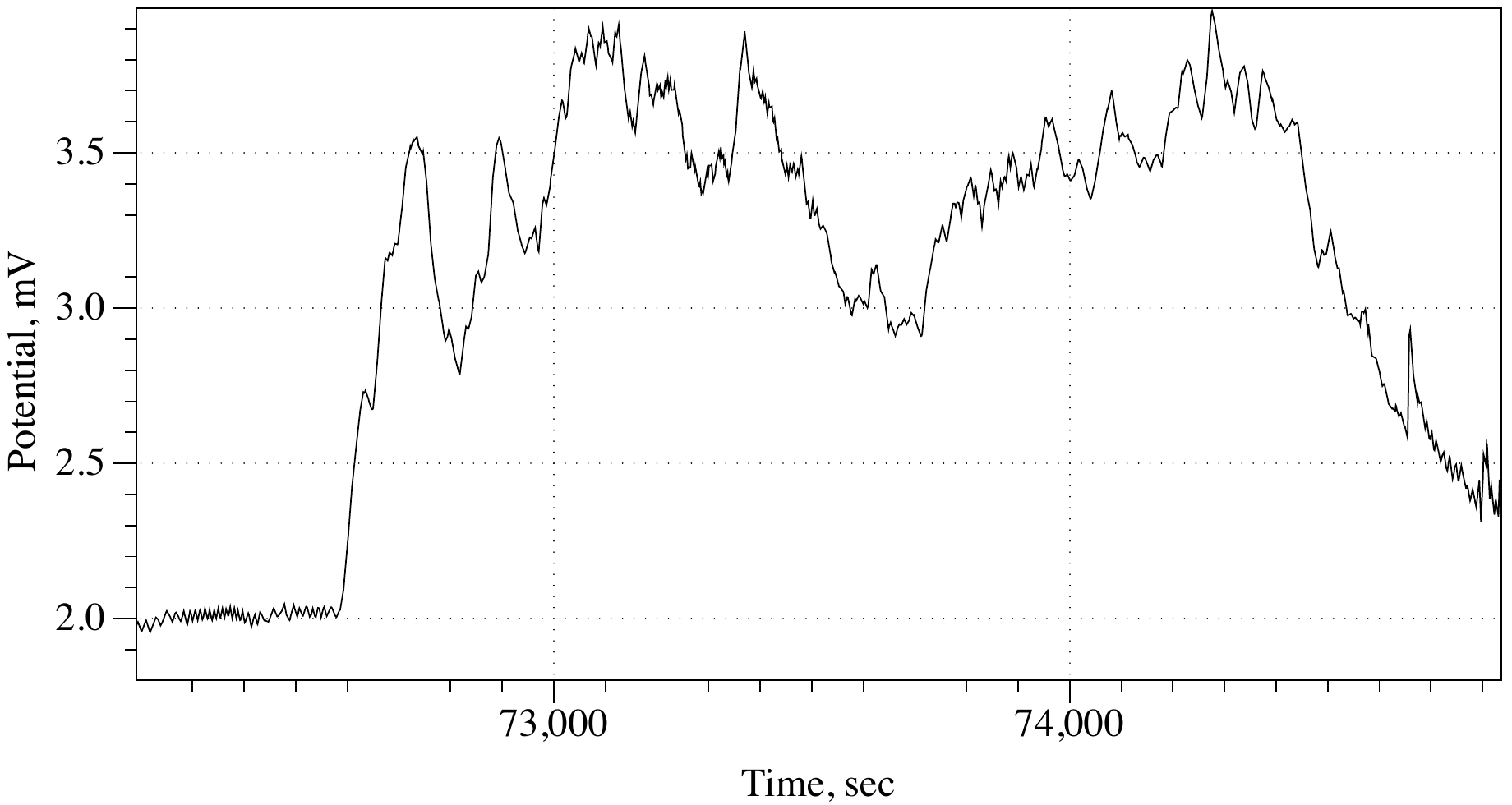}\label{fig:ResponseTo200mgweight}}
    \subfigure[]{\includegraphics[width=0.49\textwidth]{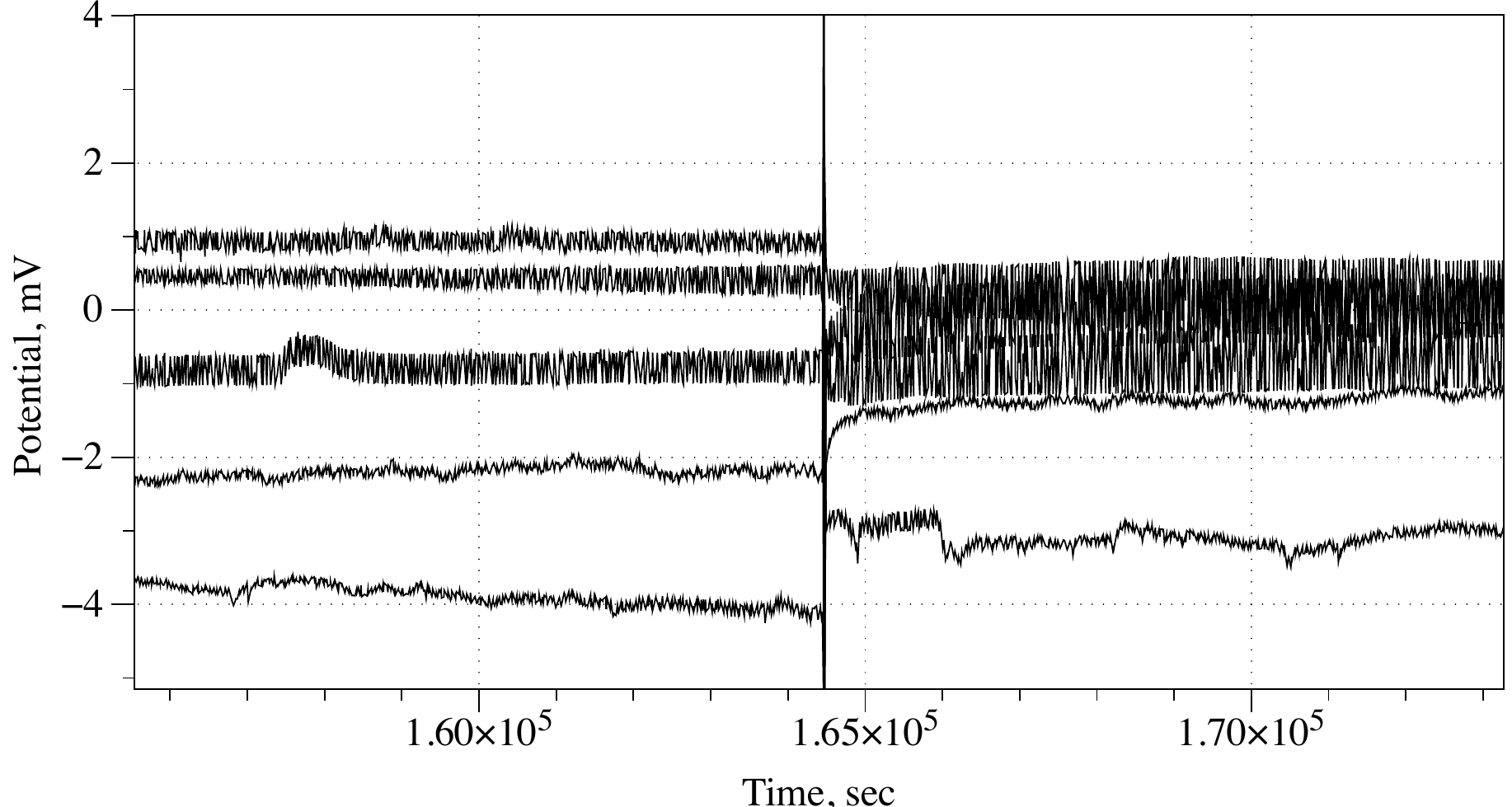}\label{fig:OverReponseToRemovalResponseTo200mgweight01}}
    \subfigure[]{\includegraphics[width=0.49\textwidth]{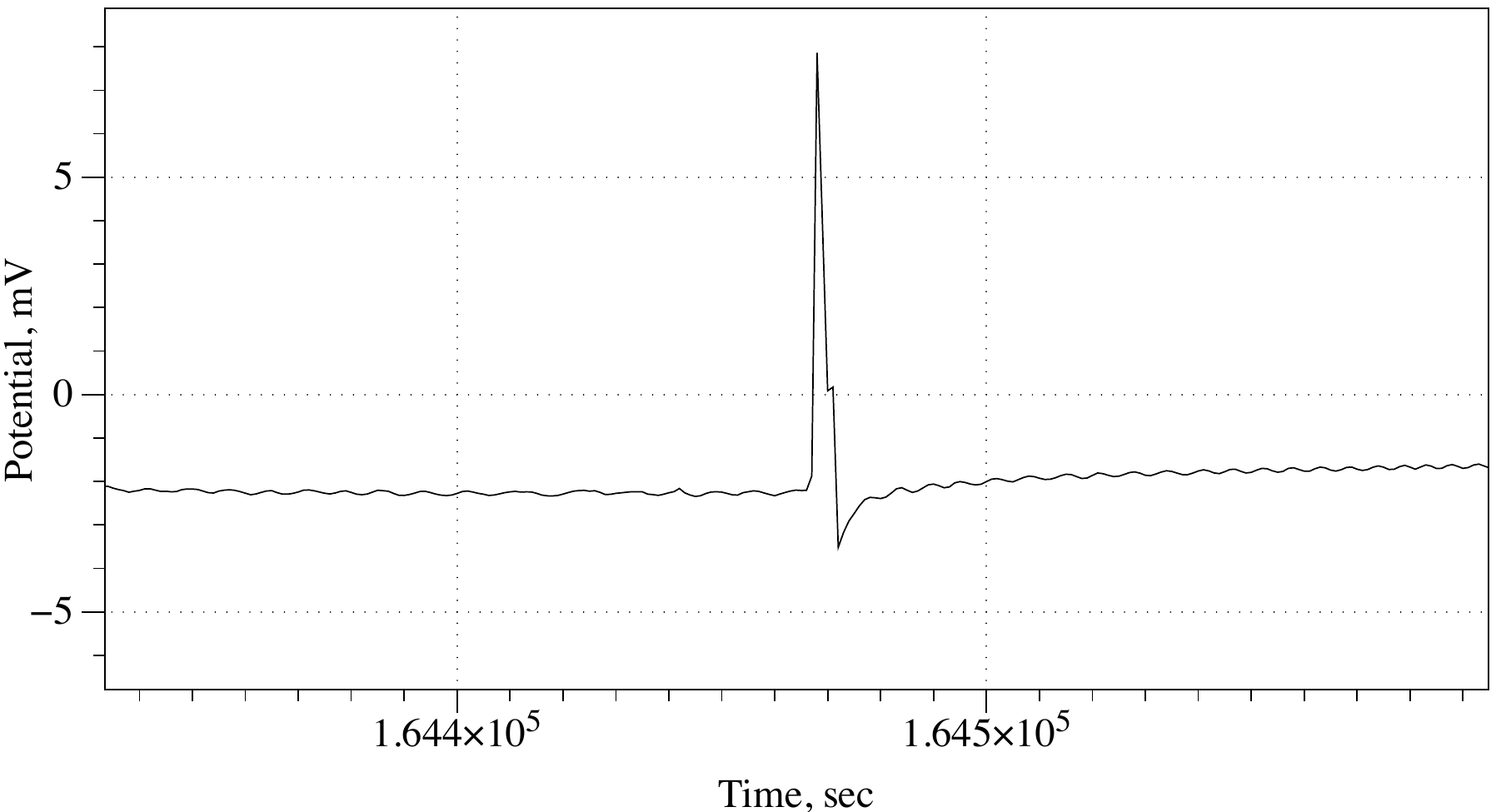}\label{fig:responseToremovalOofWeight}}
    \caption{Fungal wearable's response to stretching.
    (a)~An exemplar response to stretching of the fabric by attaching 50~g weight to it. 
    (b)~Response to removal of 200~g weight recorded from five pairs of differential electrodes.
    (c)~An exemplar response to removal of the weight recorded  on one pair of differential electrodes.
    (d)~Typical response to removal of 200~g weight, recorded on a one pair of differential electrodes.
    }
    \label{fig:my_label}
\end{figure}

To uncover the fungal wearable's response to stretching we attached 50~g and 200~g weights to the bottom part of the fabric colonised by the fungus. A typical response to the application of 50~g weight is shown in Fig.~\ref{fig:lab01_w_50_g_applied}). The response duration is 97~sec in average, $\sigma$=37~sec with average response amplitudes is 1.3~mV, $\sigma$=0.74~mV. Differential electrode pairs, labelled as channels in Fig.~\ref{fig:lab01_w_50_g_applied} have been arranged in a line from the top to the bottom (Fig.~\ref{fig:electrodes}), with Ch1 being closest to the top of the fabric and Ch13 to the bottom part. Most of the response spikes show action potential like depolarisation, repolarisation and hyperpolarisation phases. Ch1 and Ch13 show hyperpolarisation phases set up at higher, compared to that before stimulation, base potential.

On application of 200~g weight to lower part of the fabric, variety of response from differential electrodes pairs have been recorded. An exemplar response is shown in Fig.~\ref{fig:ResponseTo200mgweight}. The response has an average duration 38~min, $\sigma$=2~min, and average amplitude 1.56~mV, $\sigma$=1.24~mV. The response in the example consists of two trains of high (`high' in the frameworks of fungal temporal activity) frequency spikes. Average spike width is 80~sec ($\sigma=50$), average amplitude 0.31~mV ($\sigma$=0.32), average distance between spike in each train is 71~sec ($\sigma$=47~sec).  

Overall reaction to the removal of the stretching stimuli is in the drift of the base potentials on electrodes towards zero, e.g. in Fig.~\ref{fig:OverReponseToRemovalResponseTo200mgweight01} we see that the average based potential is -1.17~mV, $\sigma$=2~mV, before stimulation was removed, and -0.8~mV, $\sigma$=1.3~mV, after the weight was removed. A typical response, to removal of 200~g, recorded on a single Ch is shown in Fig.~\ref{fig:responseToremovalOofWeight}). The spike there has a duration of 9~sec and amplitude 11~mV.


\section{Discussion}
\label{discussion}

We demonstrated that a fabric colonised by the fungus \emph{P. ostreatus} shows distinctive sets of responses to chemical and mechanical stimulation. The response to 50~g load, Fig.~\ref{fig:lab01_w_50_g_applied}, is in the range of c.~1.5~min which might indicate that rather purely electro-mechanical events take place than reactions involving propagation of calcium waves~\cite{tuteja2007calcium}. A difference $d$ between timing of the response spikes peaks at the electrodes pairs in the line is as follows: $d$(Ch1, Ch3)=3.6~sec, $d$(Ch1, Ch5)=20~sec, $d$(Ch1, Ch7)=16~sec, $d$(Ch1, Ch13)=3~sec. This might indicate that the mycelium networks closer to the fixed end (Ch1) and the end where the load is attached (Ch13) react to the stretching first, the reaction then propagates further into the interior parts of the fabric, thus delayed reactions are recorded on the channels Ch5 and Ch7.

The response to stimulation with ethanol is in a range of 7~sec. This would rather indicate physico-chemical damages to hyphae walls and corresponding electrical responses. Would amplitude of a response spike reflect a distance to the stimulation loci? As seen in Fig.~\ref{fig:ethanolA}, on most channel the response amplitudes slightly decrease with increasing distance to the stimulation loci however more studies are required to give a certain answer. The response on the channels remote to the stimulation loci happens at the same time, as on the channel in proximity of the stimulation loci. This indicates that the response might be purely electrical (due to damage to cell walls impulse) and not due to diffusion in the fabric or volatile ethanol. 

The increase of frequency of electrical potential oscillation in a response to application of chemo-attractants or nutrients is consistent with previous studies, where intracellular electrical potential of stimulated fungi was measured~\cite{olsson1995action}. Even if in the case of malt extract solution increased spiking could be attributed to a water the experiments with dextrose, which was applied dry, show that the spiking shown increased frequency, and often amplitude, due to reaction to a chemoattractant or nutrient. The increase in amplitude of spiking five hours after the application of malt or dextrose might be due to the fungus ingesting the nutrients and transposing them across the wide mycelial network. 

In laboratory conditions the fungal wearables survived for several months being kept in high humidity conditions. When considering the fungal wearables being used in the every day life measures should be taken to preserved the moisture. For example, the fragments of fungal materials can be coated with a breathable plastic. 

In laboratory conditions the fungal wearables survive for several months being kept in high humidity conditions. For practical future applications of the fungal wearables, preserving the moisture is fundamental. For example, the fragments of fungal materials could be coated with a breathable plastic.

Future developments in the field of fungal wearables may be along the following directions. 

First direction is a computational one. We demonstrate in computational models that a fungal colony can implement a range of Boolean function~\cite{adamatzky2020boolean}. It might be possible to implement an experimental mapping between a set of stimuli and distribution of Boolean gates implemented by fungal wearables, as we demonstrated on sensing and computing organic liquid skin ~\cite{chiolerio2020tactile}. In other words, in a response to a particular stimuli the fungal wearable will generate a unique set of Boolean function.

\begin{figure}[!tbp]
    \centering
\subfigure[]{\includegraphics[width=0.35\textwidth,height=0.35\linewidth]{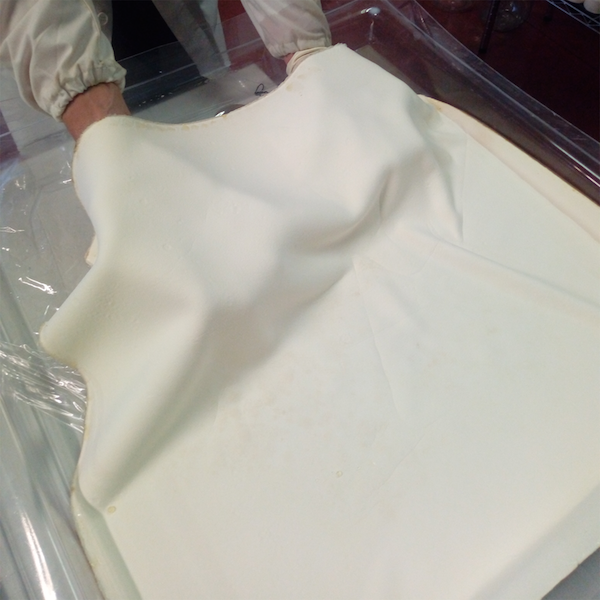}\label{fig:skinA}}
\subfigure[]{\includegraphics[width=0.35\textwidth,height=0.35\linewidth]{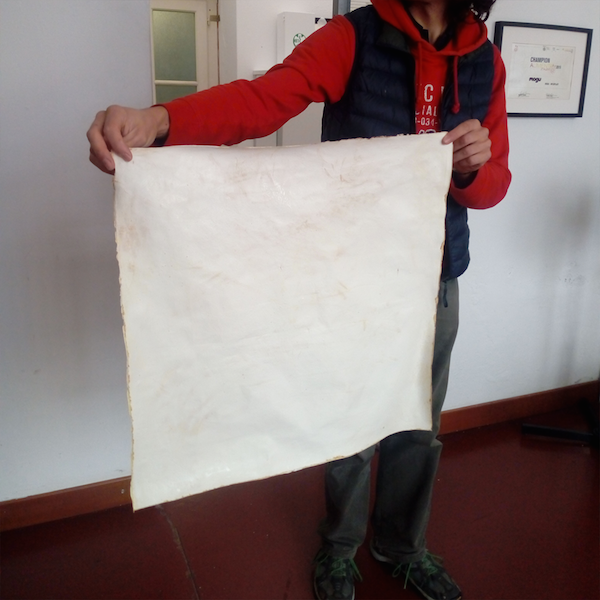}\label{fig:skinB}}
\subfigure[]{\includegraphics[width=0.35\textwidth,height=0.35\linewidth]{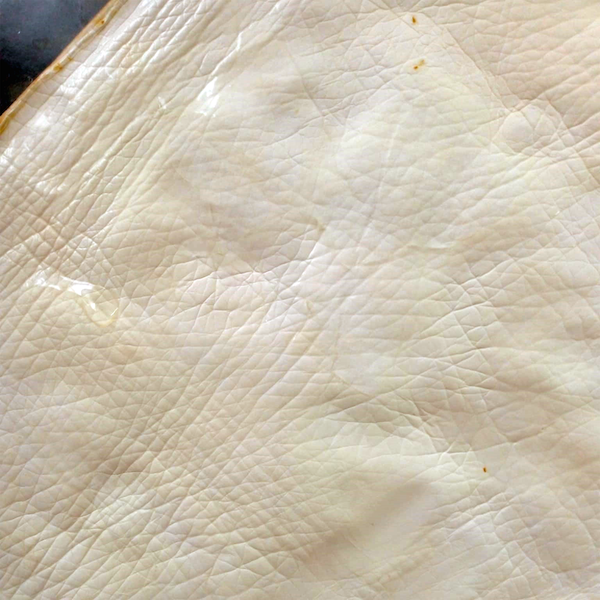}\label{fig:skinC}}
\subfigure[]{\includegraphics[width=0.35\textwidth,height=0.35\linewidth]{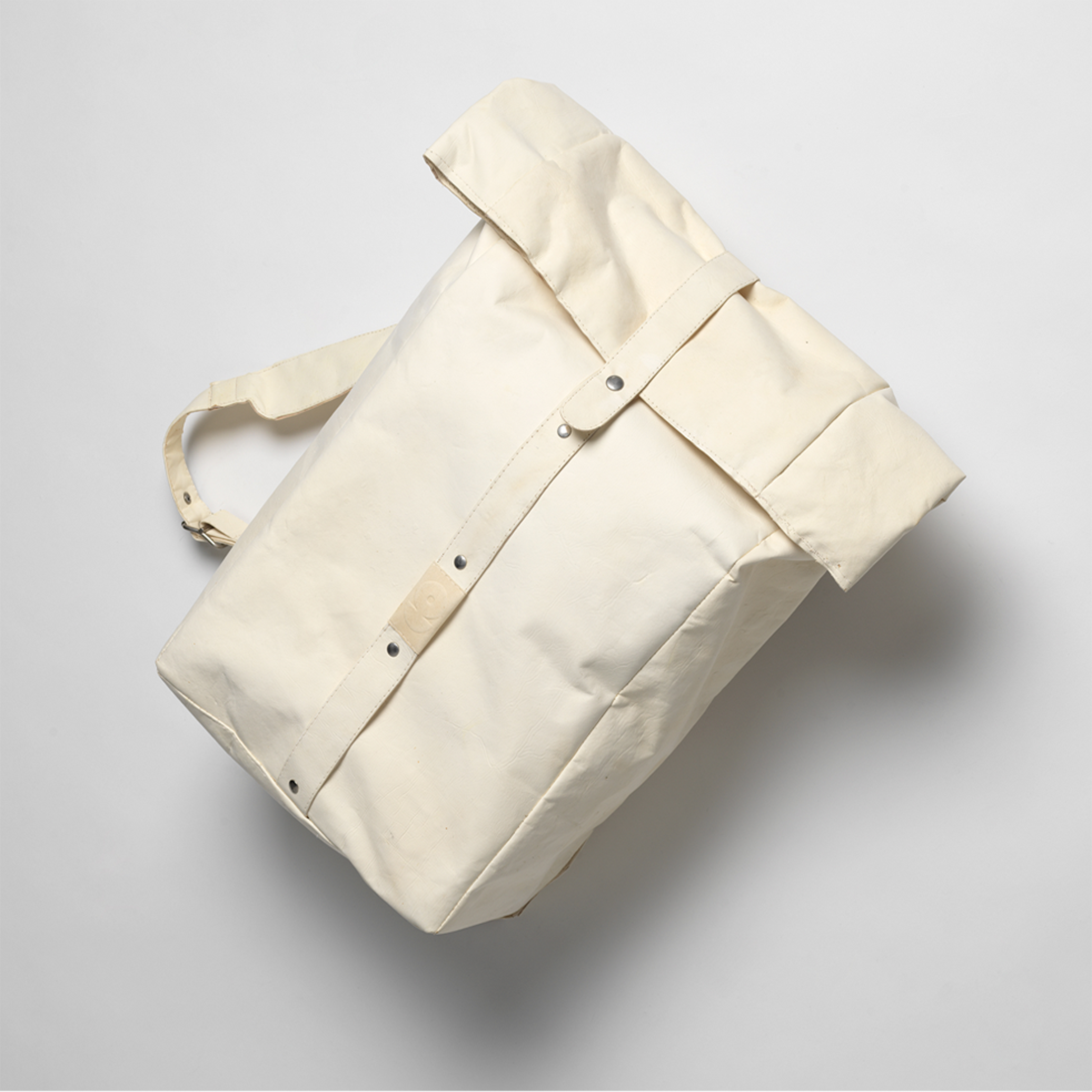}\label{fig:skinD}}
\caption{Example of fungal flexible materials grown by Mogu S.rl. and branded as PURA Flex\texttrademark. (a) Harvesting of a fungal skin, (b) size comparison with a human being, (c) texture detail resembling animal skin and (d) backpack prototype made with PURA Flex\texttrademark~material.}
    \label{fig:fungalskins}
\end{figure}

Second direction is in development of a large scale fabric made purely from mycelium --- fungal skin (Fig.~\ref{fig:fungalskins}) and tailoring the fabric into wearables. Such mycelial tissue can be prepared using trimitic polypore fungal cultures, which are apparently preferred for the production of sturdy fungal skins, such as fungal leather or \emph{mycoleather}~\cite{jones2020mycoleather}. More specifically, a fungal fabric can be prepared by pouring a homogenised slurry of a liquid culture of \emph{Ganoderma resinaceum} into a static fermentation tray and incubated for two weeks to allow the fungal hyphae to intermesh, forming a floating mat or skin~\cite{adamatzky2020fungal}. Examples of such type of fungal fabrics are shown in Fig.~\ref{fig:fungalskins}.

\begin{figure}[!tbp]
    \centering
    \subfigure[]{\includegraphics[height=0.5\linewidth]{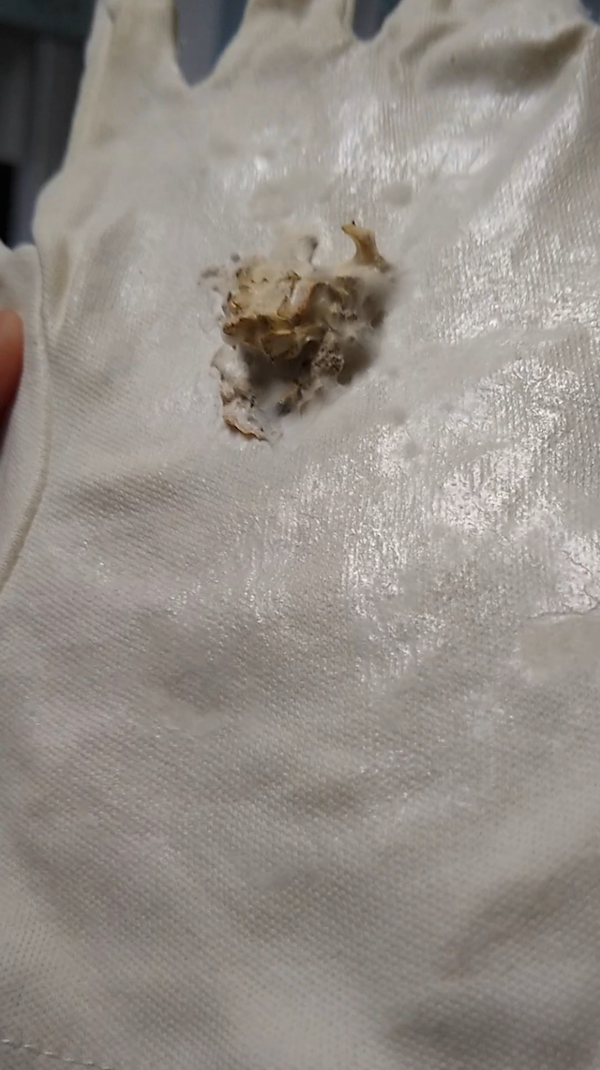}\label{fig:glove}}
    \subfigure[]{\includegraphics[height=0.5\linewidth]{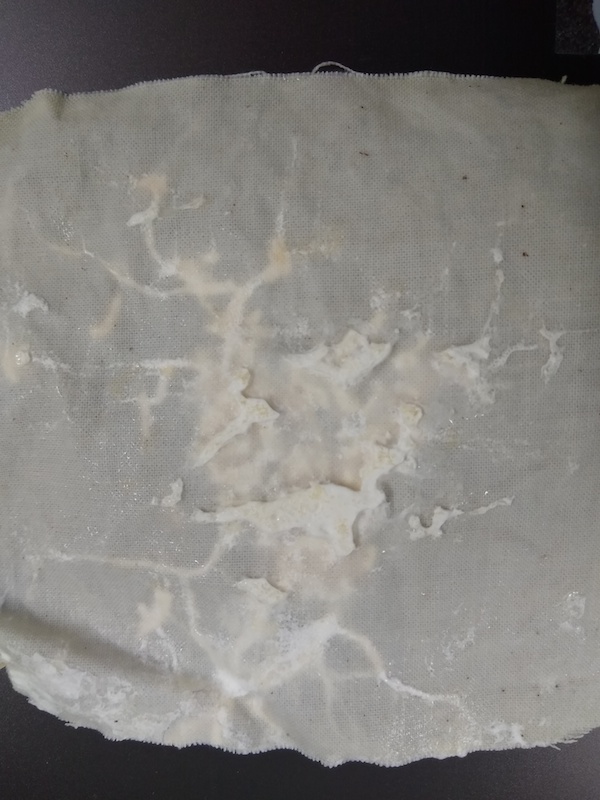}\label{fig:wires2}}
    \caption{(a)~Part of the hemp glove colonised by fungus is visible in reflected light. (b)~Stands of hyphae on the hemp fabric.}
\end{figure}

Third direction would be to culture fungi directly onto the pieces of clothing (Fig.~\ref{fig:glove}). This will allow us to achieve full response cloths and garments. 

Fourth direction in the development of fungal wearables could be in using fungal hyphae (Fig.~\ref{fig:wires2}) as wires and programmable (with e.g. light) resistor or electrically activated resistive switching devices in hybrid architectures incorporating conventional flexible electronics ~\cite{FPE2017} and live fungi. Routing the direction of the fungal wires can be done by arranging sources of attractants and repellents. Isolation of fungal wires, as well as localized connections when ordered arrays like the cross-bar array arrangement are required, could be done using inorganic materials, such as metal oxides of the proper work function deposited by means of atomic layer deposition~\cite{ALD2016}, or digitally printed over a large scale, also in case of uneven surfaces~\cite{AEM2020}.

\section*{Acknowledgement}

This project has received funding from the European Union's Horizon 2020 research and innovation programme FET OPEN ``Challenging current thinking'' under grant agreement No 858132.


\end{document}